\documentclass[twocolumn,floatfix,showpacs,pra,superscriptaddress]{revtex4}%
\usepackage{graphicx,graphics,color,epsfig}
\usepackage{bm}
\usepackage{amsmath}
\usepackage{mathrsfs}
\usepackage{amssymb}
\usepackage{subfigure}
\usepackage{amsfonts}
\usepackage{graphicx}%

\newcommand{\be}{\begin{eqnarray}}
\newcommand{\ee}{\end{eqnarray}}

\def\sgn{\mathop{\rm sgn}}

\providecommand{\U}[1]{\protect\rule{.1in}{.1in}}
\begin{document}
\title{Ring frustration and factorizable correlation functions of critical spin rings}
\author{Peng Li}
\email{lipeng@scu.edu.cn}
\affiliation{College of Physical Science and Technology, Sichuan University, 610064,
Chengdu, P. R. China}
\affiliation{Key Laboratory of High Energy Density Physics and Technology of Ministry of
Education, Sichuan University, 610064, Chengdu, P. R. China}

\author{Yan He}
\email{heyan_ctp@scu.edu.cn}
\affiliation{College of Physical Science and Technology, Sichuan University, 610064,
Chengdu, P. R. China}
\affiliation{Key Laboratory of High Energy Density Physics and Technology of Ministry of
Education, Sichuan University, 610064, Chengdu, P. R. China}

\date{\today}

\begin{abstract}
  Basing on the exactly solvable prototypical model, the critical transverse Ising ring with or without ring frustration, we establish the concept of nonlocality in a many-body system in the thermodynamic limit by defining the nonlocal factors embedded in its factorizable correlation functions. In the context of nonlocality, the valuable traditional finite-size scaling analysis is reappraised. The factorizable correlation functions of the isotropic $XY$ and the spin-1/2 Heisenberg models are also demonstrated with the emphasis on the effect of ring frustration.
\end{abstract}

\maketitle

\begin{figure}[t]
\begin{center}
\includegraphics[width=1.8in,angle=0]{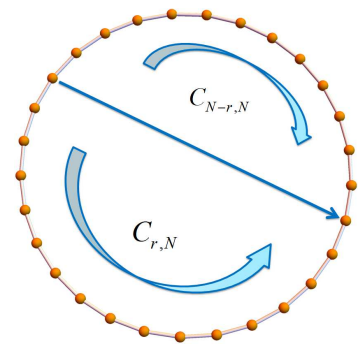}
\end{center}
\caption{(Color online) A periodic spin chain. Ring frustration occurs if the number of spins is odd and the nearest-neighbor interactions are antiferromagnetic.}%
\label{fig1}%
\end{figure}
\emph{1. Introduction.---} In quantum spin systems, highly entangled ground state can arise from geometrical frustration \cite{Diep} as well as quantum frustration \cite{Dawson}. But it is usually not easy to discern the contributions of the two different sources \cite{Giampaolo, Giampaolo2, Marzolino}. Recently, the effect of ring frustration aroused much attention due to the exotic ground state it induced \cite{Bariev, Cabrera, Campostrini, DongJSM, DongPRE, He, FranchiniEE}. A nonlocal factor in its correlation function can be extracted, which represents the pure effect of geometrical frustration \cite{DongJSM, DongPRE, He}. Ring frustration is a kind of geometrical frustration occurs for a closed chain (Fig. 1), in which no unique Ising-like state can prevail in the ground state and minimize the system's energy alone. Unlike the usual local geometric frustration on the triangular or Kagom\'{e} lattices, the ring frustration is of a nonlocal nature in that: (i) One must walk all the way round the ring to make sure of the presence of spin frustration, i.e., the frustration is somewhat \emph{weak} \cite{FranchiniEE}. (ii) It may significantly change the bulk property of the low-energy states \cite{DongJSM, DongPRE}.

On the other hand, the concept of thermodynamic limit resides in the central part of statistical mechanics, with which the critical phenomena must associated. In theoretical calculations, we manage to match the physical systems of Avogadro's number of spins by setting the number of spins in the models to a mathematical infinity, $N\rightarrow\infty$. In traditional treatment, we often take the limit, $N\rightarrow\infty$, at the very beginning stage of calculations, which facilitates us to employ useful transforms, such as the substitution of the sum of momentum number $q$ with an integral (in $D$ dimensions), $(1/N)\sum_{q}[\cdots]=\int\mathrm{d}^{D}q/(2\pi)^D[\cdots]$, to work out desired quantities. Thus $N$ will disappear in the final results. For example, critical spin chains have been found to exhibit algebraically decaying correlation functions like
\begin{equation}
C_{r,\infty}\sim\frac{b}{r^\eta},
\label{algebraic}
\end{equation}
where $b$ and $\eta$ are real numbers.

Can we defer the setting of the limit, $N\rightarrow\infty$, till the end of calculation? And if so, what can we get from it? In this work, we shall demonstrate that the concepts of locality and nonlocality can be well distinguished and defined for a ring system in the limit, $N\rightarrow\infty$. We establish the full framework for extracting the nonlocal factors in the correlation function basing on an exactly solvable prototype, the transverse Ising ring at its phase transition point. Then we reappraise the usefulness of the finite-size scaling analysis in this framework and apply it to the isotropic $XY$ and Heisenberg rings with emphasis on the effect of ring frustration.

\emph{2. Definitions of nonlocal factors.---} Suppose the total number of spins on a ring can approach the limit, $N\rightarrow\infty$ (Fig. \ref{fig1}). We concern the spin correlation function of the ground state $|E_{0}\rangle$,
\begin{align}
C_{r,N}=\langle E_{0}|\sigma_{j}^{a}\sigma_{j+r}^{a}|E_{0}\rangle,
\label{Correlator}
\end{align}
where $\sigma_{j}^{a}$ ($a=x,y,z$) are Pauli matrices (we shall only consider $a=x$ for abbreviation). Obviously, a cyclic relation holds for the correlation function of two spins with distance $r$,
\begin{equation}
C_{r,N}=C_{N-r,N}.
\label{cyclic}
\end{equation}

The main idea is that the results may be different if the limit, $N\rightarrow\infty$, is made at two different occasions:

(i) If setting $N\rightarrow\infty$ at the beginning stage of calculation, we denote the result with the function
\begin{align}
C_{r,\infty}\equiv C_{r,\lim_{N\rightarrow\infty}N}.
\end{align}
The example in Eq. (\ref{algebraic}) falls into this case.

(ii) If setting $N\rightarrow\infty$ at the end of calculation, we get
\begin{align}
C^{(O)}(\alpha)&\equiv\lim_{L\rightarrow\infty}C_{r,2L+1}, \label{C(O)}\\
C^{(E)}(\alpha)&\equiv\lim_{L\rightarrow\infty}C_{r,2L}, \label{C(E)}
\end{align}
for $N=2L+1\in Odd$ and $N=2L\in Even$ respectively, where we have defined a parameter
\begin{equation}
\alpha=\lim_{N\rightarrow\infty}\frac{r}{N}.
\label{alpha}
\end{equation}
Its value can be restricted to the range $0\leq\alpha<1/2$ due to the ring geometry. It is natural to put the distances into three categories in the limit, $N\rightarrow\infty$:

(i) The distance is \emph{local} if $r\approx 1$.

(ii) The distance is \emph{near local} if $r\gg 1$ and $\alpha=0$, just like Eq. (\ref{algebraic}).

(iii) The distance is \emph{nonlocal} if $\alpha\neq 0$.

In this work, we present a clear prototype, the transverse Ising ring at its phase transition point, to demonstrate the differences among $C_{r,\infty}$, $C^{(O)}(\alpha)$, and $C^{(E)}(\alpha)$. More important, we propose three nonlocal factors defined as ratios. The first two of them are for the measure of nonlocality for $N=2L+1\in Odd$ and $N=2L\in Even$ respectively,
\begin{align}
R^{(O)}(\alpha)&=\frac{C^{(O)}(\alpha)}{C_{r,\infty}},\label{R(O)}\\
R^{(E)}(\alpha)&=\frac{C^{(E)}(\alpha)}{C_{r,\infty}},\label{R(E)}
\end{align}
and the third is for the measure of the effect of ring frustration,
\begin{align}
R(\alpha)&=\frac{R^{(O)}(\alpha)}{R^{(E)}(\alpha)}.
\label{R}
\end{align}
These definitions hold only if the correlation functions, $C^{(O)}(\alpha)$ and $C^{(E)}(\alpha)$, are factorizable.

\emph{3. Prototype: Transverse Ising ring at its phase transition point.---} The transverse Ising model is a special case of the general Hamiltonian,
\begin{equation}
H(\gamma,h)=\sum_{j=1}^{N}\left(\frac{1+\gamma}{2}\sigma_{j}^{x}\sigma_{j+1}^{x}+
\frac{1-\gamma}{2}\sigma_{j}^{y}\sigma_{j+1}^{y}\right)-h\sum_{j=1}^{N}\sigma_{j}^{z},
\label{XY}
\end{equation}
where $\gamma$ and $h$ are parameters for anisotropy and transverse field. This general Hamiltonian can be faithfully solved in the framework of $a$-cycle problem \cite{Lieb, Mazur, DongMPLB}.

Let us focus on the transverse Ising ring at its phase transition point,
\begin{equation}
H^{\mathrm{TI}}=H(1,1).
\label{TI}
\end{equation}
It is direct to work out the two-point longitudinal correlation function with the aid of Jordan-Wigner fermion representation \cite{J-W} (please see details in Appendix A).

For $N=2L\in Even$, we get the correlation function in the form of Toeplitz determinant,
\begin{align}
C_{r,N}=\left(-\frac{1}{N}\right)^{r}\det\Big[\csc\frac{(\mu_j+\nu_k)\pi}{2N}\Big]_{0\leq j,k\leq r-1},
\label{Cr2L_TI}%
\end{align}
where $\mu_j=2j+1,~~~\nu_k=-2k$. While for $N=2L+1\in Odd$, we get
\begin{align}
C_{r,N}=\left(\frac{1}{N}\right)^{r}\det\Big[1-\cot\frac{\left(\mu_j+\nu_k\right)\pi}{2N}\Big]_{0\leq j,k\leq r-1}.
\label{Cr2L1_TI}%
\end{align}
At this moment, if we set $N\rightarrow\infty$, both Eqs. (\ref{Cr2L_TI}) and (\ref{Cr2L1_TI}) will become the same Cauchy determinant that can be worked out and leads to the well-known asymptotic formula \cite{Wu, McCoyPR, McCoy},
\begin{align}
C_{r,\infty}&=\left(\frac{2}{\pi}\right)^{r}\det\Big[\frac{1}{\mu_j+\nu_k}\Big]_{0\leq j,k\leq r-1}\nonumber\\
&=\left(\frac{2}{\pi}\right)^{r}\frac{\prod_{0\leq j<k\leq r-1}(\mu_j-\mu_k)(\nu_j-\nu_k)}{\prod_{j=0}^{r-1}\prod_{k=0}^{r-1}(\mu_j+\nu_k)}\nonumber\\
&\approx (-1)^r\frac{b_1}{r^{1/4}},\label{C_TTWu}
\end{align}
where $b_1 =\mathrm{e}^{1/4}2^{1/12}A^{-3}\approx 0.645002448$, $A$ is the Glaisher's constant.

To work out the determinants in Eqs. (\ref{Cr2L_TI}) and (\ref{Cr2L1_TI}) rigorously, we employ the identities,
\begin{align}
\det&\Big[\frac{1}{\sin(a_i+b_j)}\Big]_{0\leq i,j\leq n-1} \nonumber\\
&=\frac{\prod_{0\leq i<j\leq n-1}\sin(a_i-a_j)\sin(b_i-b_j)}{\prod_{0\leq i,j\leq n-1}\sin(a_i+b_j)},\label{Pn}
\\
\det&\Big[\frac{\cos(a_i+b_j+\phi)}{\sin(a_i+b_j)}\Big]_{0\leq i,j\leq n-1}\nonumber\\
&=\frac{\prod_{0\leq i<j\leq n-1}\sin(a_i-a_j)\sin(b_i-b_j)}{\prod_{0\leq i,j\leq n-1}\sin(a_i+b_j)}\nonumber\\
&\hspace{1em}\times\cos\Big[\sum_{i=0}^{n-1}(a_i+b_i)+\phi\Big]\cos^{n-1}\phi,
\label{Qn}
\end{align}
that can be proved by mathematical recursion (Appendix B). By Eq. (\ref{Pn}) and (\ref{Qn}), we get
\begin{align}
C_{r,2L}&=(-1)^{r}S_{r,2L}   \label{Cr2L}
\end{align}
and
\begin{align}
C_{r,2L+1}&=(-1)^{r}S_{r,2L+1}B_{1}(\alpha)  \label{Cr2L1}%
\end{align}
respectively, where
\begin{align}
&S_{r,N}=\frac{\prod_{0\leq j<k\leq r-1}\sin\frac{(\mu_j-\mu_k)\pi}{2N}\sin\frac{(\nu_j-\nu_k)\pi}{2N}}{N^{r}\prod_{j=0}^{r-1}\prod_{k=0}^{r-1}\sin\frac{(\mu_j+\nu_k)\pi}{2N}},
\label{SrN0}\\
&B_{1}(\alpha)=\cos\frac{\alpha \pi}{2}-\sin\frac{\alpha \pi}{2}.
\end{align}

Although Eqs. (\ref{Cr2L}) and (\ref{Cr2L1}) are rigorous for arbitrary $N$ and $r$, but they are not convenient to tell whether they are factorizable in the limit $N\rightarrow\infty$ and $r\rightarrow\infty$. Denoting $\theta=\frac{\pi}{2N}$, we transform $S_{r,N}$ in Eq. (\ref{SrN0}) to
\begin{align}
S_{r,N}=\frac{\prod_{1\leq m\leq r-1}\left(\cos^{2}\theta-\cot^{2}m\theta\sin^{2}\theta\right)^{m-r}}{\left(N \sin\theta\right)^{r}}.
\label{SrN1}
\end{align}
Then, noticing the identity (for arbitrary $N=2L+1$),
\begin{align}
\frac{1}{N \sin\theta}=\prod_{1\leq m\leq L}\left(\cos^{2}\theta-\cot^{2}m\theta\sin^{2}\theta\right),
\label{identitySrN}
\end{align}
we find that
\begin{align}
\ln S_{r,N}&=\sum_{m=1}^{r-1}m \ln\left(\cos^{2}\theta-\cot^{2}m\theta\sin^{2}\theta\right) \nonumber\\
&+r \sum_{m=r}^{(N-1)/2}\ln\left(\cos^{2}\theta-\cot^{2}m\theta\sin^{2}\theta\right).
\label{lnSrN0}
\end{align}
Next, by substituting the Taylor expansion,
\begin{align}
\ln &\left(\cos^{2}\theta -\cot^{2}m\theta\sin^{2}\theta\right)\nonumber\\
&=\ln(1-\frac{1}{4m^2})-\frac{1}{3}\theta^2-\frac{1+24m^2}{90}\theta^4-\cdots,
\end{align}
into Eq. (\ref{lnSrN0}) and accomplishing the summations with the index $m$, we arrive at
\begin{align}
\ln S_{r,N}=-\frac{1}{4}\ln r+\ln b_{1}+h(\alpha)+O(\frac{1}{N}),
\label{lnSrN1}
\end{align}
where $h(\alpha)$ is a sum containing two convergent expansions (for more terms, please see Appendix B)
\begin{align}
h(\alpha)&=\frac{\alpha}{2}-(\frac{\pi^{2}\alpha^{2}}{24}+\frac{\pi^{4}\alpha^{4}}{240}+\cdots)\nonumber\\
&-\Big[\frac{\pi^2 \alpha(1-2\alpha)}{24}+\frac{\pi^4 \alpha(1-8\alpha^3)}{1440}+\cdots\Big].
\label{}
\end{align}
At this last moment, we are able to keep the parameter $\alpha$ after ignoring the terms in order of $O(\frac{1}{N})$, and get
\begin{align}
S(\alpha)\equiv\lim_{N\rightarrow\infty}S_{r,N}=\frac{b_{1}}{r^{1/4}}\mathrm{e}^{h(\alpha)}.
\end{align}
Now we can reap the accurate nonlocal factor,
\begin{align}
R^{(O)}(\alpha)=\mathrm{e}^{h(\alpha)}B_{1}(\alpha).
\label{R(O)_TIC}
\end{align}
And due to the small difference ($N=2L+1$),
\begin{align}
\max(S_{r,N}-S_{r,N-1})\sim \frac{1}{N^{5/4}}\rightarrow 0,
\label{dS}
\end{align}
the above calculation is also true for $N=2L\rightarrow\infty$. Thus we get the other two nonlocal factors
\begin{align}
&R^{(E)}(\alpha)=\mathrm{e}^{h(\alpha)},
\label{R(E)_TIC}\\
&R(\alpha)=B_{1}(\alpha).
\label{R_TIC}
\end{align}
The nonlocal factors are illustrated in Fig. 2. We see that $R^{(O)}(\alpha)$ and $R(\alpha)$ are quite close since $R^{(E)}(\alpha)$ deviates not far from 1. As a compatible result, previous studies revealed a nonlocal factor, $R(\alpha)=1-2\alpha$, for the kink phase of the transverse Ising chain with ring frustration \cite{Campostrini,DongJSM}, and now we may deem the model exhibits the other two trivial factors, $R^{(E)}(\alpha)=1$ and $R^{(O)}(\alpha)=R(\alpha)$.

\begin{figure}[t]
\begin{center}
\includegraphics[width=3.25in,angle=0]{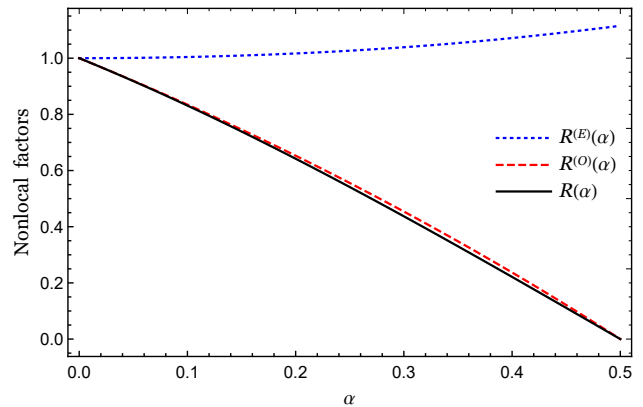}
\end{center}
\caption{(Color online) Nonlocal factors of the transverse Ising ring at its phase transition point.}%
\label{fig2}%
\end{figure}

\emph{4. Nonlocal factors and finite-size scaling.---} However, many models in the limit $N\rightarrow\infty$ can not be solved as so exactly as the transverse Ising ring at its phase transition point. So, instead of Eqs. (\ref{R(O)})-(\ref{R}), we have to conjecture the trends of the finite-size version of the ratios,
\begin{align}
R^{(O)}_{r,2L+1}=\frac{C_{r,2L+1}}{C_{r,\infty}}&~~\longrightarrow~~R^{(O)}(\alpha),\label{R(O)rN}\\
R^{(E)}_{r,2L}=\frac{C_{r,2L}}{C_{r,\infty}}&~~\longrightarrow~~R^{(E)}(\alpha),\label{R(E)rN}\\
R_{r,2L+1}=\frac{R^{(O)}_{r,2L+1}}{R^{(E)}_{r,2L}}&~~\longrightarrow~~R(\alpha), \label{RrN}
\end{align}
with the system's size increasing. This is nothing but the famous finite-size scaling (FSS) hypothesis. In fact, as a scaling function, $R^{(E)}_{r,2L}$ has been studied tremendously by numerical methods for many models in the past decades \cite{Kaplan}. While the other two, $R^{(O)}_{r,2L+1}$ and $R_{r,2L+1}$, have been somewhat overlooked so far, till the effect of ring frustration makes them prominent \cite{DongJSM}. And as one of the most important inferences, the scaling function observed in the FSS analysis may truly approximate the nonlocal factor of an infinite system. This conclusion brings a wonderful reappraisal for FSS in that it is a valuable method for exploring nonlocality in many-body systems. We address this by figuring out the nonlocal factors of the isotropic $XY$ and spin-1/2 Heisenberg rings.

\emph{5. Isotropic $XY$ ring.---} Now we turn to the isotropic $XY$ model,
\begin{equation}
H^{XY}=H(0,0).
\label{XY}
\end{equation}
The solutions of its $a$-cycle problem is quite delicate \cite{Lieb, DongJSM, DongMPLB}. It turns out that the systems with $N=4K, 4K+2\in Even$ and $N=4K+1, 4K+3\in Odd$ should be solved separately. For simplicity and without loss of generality, we present the numerical results of $N=4K\in Even$ and $N=4K+1\in Odd$ here (for more information, please see Appendix A).

For $N=4K\in Even$ the ground state is unique and the excitations are gapless. The correlation function is expressed by a Toeplitz determinant,
\begin{align}
C_{r,N}=\det\Big[\mathscr{T}_{j-k,N}\Big]_{1\leq j,k\leq r},
\label{CrN4K_XY}%
\end{align}
where the element reads
\begin{align}
\mathscr{T}_{n,N}=\left\{
\begin{array}[c]{ll}%
0,&(n=1);\\
-\frac{2}{N}\csc\frac{\pi (n-1)}{N}\sin\frac{\pi (n-1)}{2},&(\mathrm{other~} n).%
\end{array}
\right.
\label{scrTrN4K}
\end{align}
Again, at this moment, if setting the limit, $N\rightarrow\infty$, before the evaluation of the Toeplitz determinants, the element in Eq. (\ref{scrTrN4K}) becomes the same one obtained originally by Lieb \emph{et al.} \cite{Lieb},
\begin{align}
\mathscr{T}_{n,\infty}\approx\left\{
\begin{array}[c]{ll}%
0,~~&(n\in odd);\\
\frac{2}{\pi (n-1)}\cos\frac{n \pi}{2},~~&(n\in even).
\end{array}
\right.
\label{}
\end{align}
Basing on it, McCoy found an asymptotic formula \cite{McCoyPR},
\begin{align}
C_{r,\infty}\approx(-1)^r\frac{b_{2}}{r^{1/2}},
\label{Crinf_xy}
\end{align}
where $b_{2} =\mathrm{e}^{1/2}2^{2/3}A^{-6}\approx 0.588352664$. It is easy to verify the original observation by Kaplan \emph{et al.} that the numerical result of Eq. (\ref{CrN4K_XY}) deviates from Eq. (\ref{Crinf_xy}) by a factor \cite{Kaplan}, 
\begin{align}
R^{(E)}(\alpha)=1+0.28822\sinh^{2}(1.673\alpha).
\label{R(E)_XY}
\end{align}
This factor was ascribed to the finite-size effect. Now in the context of nonlocality, we can reasonably say it truly reflects the nonlocal property when the system's size approaches infinity.

\begin{figure}[t]
\begin{center}
\includegraphics[width=3.25in,angle=0]{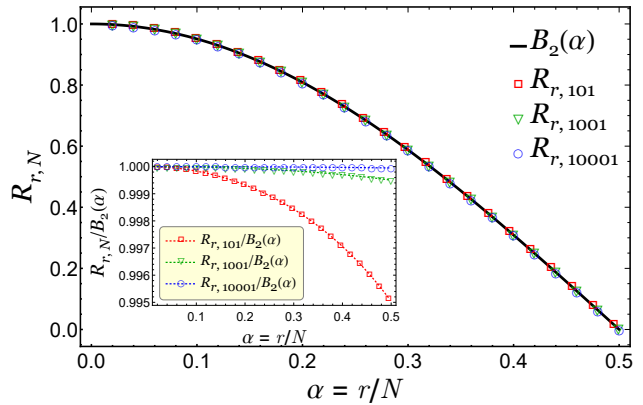}
\end{center}
\caption{(Color online) $R_{r,N}$ with $N=101, 1001$ and $10001$ for the isotropic $XY$ model. The data collapse to the proposed scaling
curve $B_{2}(\alpha)$ very accurately. The ratios, $R_{r,N}/B_{2}(\alpha)$, in the inset demonstrate how the data approach the curve $B_{2}(\alpha)$ with $N$ increasing.}%
\label{fig3}%
\end{figure}

While for $N=4K+1\in Odd$, there are four degenerate ground states. Without loss of generality, we deduce the correlation function for one of them as (please see Appendix A for more details)
\begin{align}
C_{r,N}=\det\Big[\mathscr{T}_{j-k,N}+\frac{2\beta_{Q_{o}}}{N}\mathrm{e}^{\mathrm{i}(j-k)Q_{o}}\Big]_{1\leq j,k\leq r},
\label{Cr2L1_XY}%
\end{align}
where $\beta_{Q_{o}}=\sgn(\cos Q_{o})\operatorname{e}^{-\operatorname*{i}Q_{o}}$, $Q_{o} = \frac{N-1}{2N}\pi$, and
\begin{align}
\mathscr{T}_{n,N}=\left\{
\begin{array}[c]{ll}%
-\frac{1}{N},~~&(n =1);\\
-\frac{2}{N}\csc\frac{(n-1)\pi}{N}\sin\frac{(1+N)(n-1)\pi}{2N}, ~&(\mathrm{other~}n).%
\end{array}
\right.
\label{scrTr4K1}
\end{align}
We directly work out the data of $R_{r,N}$ with $N=101,1001,10001$ according to Eq. (\ref{RrN}). We found the data perfectly collapse to the curve (Fig. 3)
\begin{align}
B_{2}(\alpha)=\left(\cos\frac{\alpha \pi}{2}\right)^{2}-\left(\sin\frac{\alpha \pi}{2}\right)^{2},
\label{B2}
\end{align}
which suggests the nonlocal factor due to pure ring frustration is
\begin{align}
R(\alpha)=\lim_{N\rightarrow\infty}R_{r,N}= B_{2}(\alpha),
\label{R_XY}
\end{align}
The nonlocal factor $R^{(O)}(\alpha)$ can be inferred from Eqs. (\ref{R(E)_XY}) and (\ref{R_XY}) easily.

\begin{figure}[t]
\begin{center}
\includegraphics[width=3.3in,angle=0]{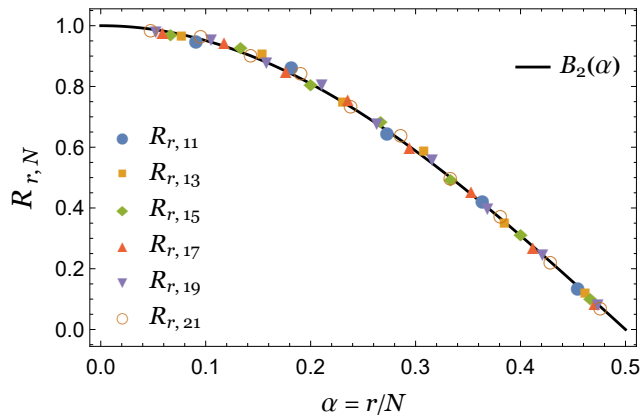}
\end{center}
\caption{(Color online) $R_{r,N}$ with $N$ from $11$ to $21$ for the spin-1/2 Heisenberg model. The data collapse to the proposed scaling curve $B_{2}(\alpha)$ very well. }%
\label{fig4}%
\end{figure}

\emph{6. Spin-1/2 Heisenberg ring.---} We solve the spin-1/2 Heisenberg ring
\be
H^{H}=J\sum_{i=1}^N(\sigma^x_i \sigma^x_{i+1}+\sigma^y_i \sigma^y_{i+1}+\sigma^z_i \sigma^z_{i+1})
\ee
by Bethe ansatz \cite{Karbach, Takahashi, FranchiniBook} and get the data of the correlations $C_{r,N}$ with $N$ from $N=10$ to $21$ (please see details in Appendix C). Then the data for $R_{r,N}$ are produced according to Eq. (\ref{RrN}). The result is shown in Fig. 4. We see the data also collapse to the curve $B_{2}(\alpha)$ quite well, which suggests it shares the same nonlocal factor, $R(\alpha)= B_{2}(\alpha)$, with the $XY$ ring. As for the nonlocal factor $R^{(E)}(\alpha)$, one can refer to the proposal by Hallberg \emph{et al.} \cite{Hallberg}, $R^{(E)}(\alpha)=[1+0.28822\sinh^{2}(1.673\alpha)]^{1.805}$, from which $R^{(O)}(\alpha)$ is easily inferred.

\emph{7. Conclusion and discussion.---} In brief, we have proposed a well-defined concept of nonlocality in the infinite spin rings. Three popular critical spin models are presented as cases in point. The transverse Ising ring serves as a prototype since it is exactly solvable. Basing on it, we establish the framework for extracting the nonlocal factors in the correlation functions with emphasis on the effect of ring frustration. This prototype let us to see clear the essential role of traditional FSS analysis in the calculations of these spin models. The usefulness of FSS analysis in the framework is demonstrated by numerical solutions of the nonlocal factors in the correlation functions of isotropic $XY$ and spin-1/2 Heisenberg rings.

In previous studies, the odevity of the total number of spins, $N$, has not been seriously handled. When $N\in even\rightarrow\infty$, the existence of the nonlocal factor, $R^{(E)}(\alpha)$, suggests $C_{r,\infty}$ loses some quantum entanglement information that a true periodic and infinite system ought to possess \cite{Dawson}. While when $N\in odd\rightarrow\infty$, i.e. when ring frustration is at presence, the system's bulk properties may be changed dramatically, thus the discrepancy between $C^{(O)}(\alpha)$ and $C_{r,\infty}$ becomes significantly large \cite{DongJSM, DongPRE}. Although $C^{(O)}(\alpha)$ contains both geometrical and quantum frustrations \cite{Dawson, Giampaolo}, our conclusion is that $C^{(O)}(\alpha)$ is factorizable in the limit $N\rightarrow\infty$,
\begin{equation}
C^{(O)}(\alpha)=R(\alpha)C^{(E)}(\alpha)=R(\alpha)R^{(E)}(\alpha)C_{r,\infty},
\label{factorizable}
\end{equation}
in which $R(\alpha)$ denotes the part of pure geometrical frustration.

We thank Jian-Jun Dong for useful discussions. This work is supported by NSFC under Grants no.

\newpage

\newpage

\begin{widetext}

\appendix

\section{The $\Huge{a}$-cycle problem for the $XY$ ring}

\subsection{General formulae}

The general Hamiltonian for the anisotropic $XY$ model reads,
\begin{equation}
H(\gamma,h)=\sum_{j=1}^{N}\left(\frac{1+\gamma}{2}\sigma_{j}^{x}\sigma_{j+1}^{x}+
\frac{1-\gamma}{2}\sigma_{j}^{y}\sigma_{j+1}^{y}\right)-h\sum_{j=1}^{N}\sigma_{j}^{z},
\label{XY}
\end{equation}
where $\gamma$ and $h$ are parameters for anisotropy and transverse field.

We consider a periodic boundary condition, $\sigma_{j}^{a}=\sigma_{j+N}^{a}$, which leads to a so-called $a$-cycle problem \cite{Lieb,Mazur} after the spins, $\sigma_{j}^{a}$, are mapped to the Jordan-Wigner fermions, $c_{j}$ and $c_{j}^{\dag}$. The Jordan-Wigner transformation reads \cite{J-W},
\begin{equation}
\sigma_{j}^{+}=\left(  \sigma_{j}^{x}+\operatorname{i}\sigma_{j}^{y}\right)
/2=c_{j}^{\dag}\exp(\operatorname{i}\pi\sum_{l<j}c_{l}^{\dag}c_{l}).
\end{equation}
Then, by the means of Fourier transformation
\begin{equation}
c_{q}=\frac{1}{\sqrt{N}}\sum_{j=1}^{N}c_{j}\exp\left(  \operatorname{i}q\,j\right),
\end{equation}
and Bogoliubov transformation (we adopt the same notations as that in Ref. \cite{DongJSM}),
\begin{equation}
\eta_{q}=u_{q}c_{q}-\operatorname{i}v_{q}c_{-q}^{\dagger},~~(q\neq 0,\pi)
\end{equation}
with
\begin{align}
u_{q}^{2}=\frac{1}{2}&\left(  1+\frac{\epsilon(q)}{\omega(q)}\right)
,v_{q}^{2}=\frac{1}{2}\left(  1-\frac{\epsilon(q)}{\omega(q)}\right)
,2u_{q}v_{q}  =\frac{\Delta(q)}{\omega(q)},\nonumber\\
&\epsilon(q)=\cos{q}-h,\Delta(q)=\gamma\sin{q},\nonumber\\
&\omega(q)=\sqrt{\epsilon(q)^{2}+\Delta(q)^{2}},
\end{align}
we diagonalize the Hamiltonian with the number of lattice sites $N\in Even(E)$ or $N\in Odd(O)$ in the $even(e)$ or $odd(o)$ channels as (so there are four combinations),
\begin{align}
&H^{(E,o)}=\epsilon(0)( 2c_{0}^{\dagger}c_{0}-1)+\epsilon(\pi)( 2c_{\pi}^{\dagger}c_{\pi}-1) +\sum_{q\in q^{(E,o)},q\neq0,\pi}\omega(q)\left(  2\eta_{q}^{\dagger}\eta_{q}-1\right)  ,\label{H(E,o)}\\
&H^{(E,e)}=\sum_{q\in q^{(E,e)}}\omega(q)\left(  2\eta_{q}^{\dagger}\eta_{q}-1\right)  , \label{H(E,e)}\\
&H^{(O,o)}=\epsilon(0)( 2c_{0}^{\dagger}c_{0}-1) +\sum_{q\in q^{(O,o)},q\neq0}\omega(q)\left(  2\eta_{q}^{\dagger}\eta_{q}-1\right)  ,\label{H(O,o)}\\
&H^{(O,e)}=\epsilon(\pi)( 2c_{\pi}^{\dagger}c_{\pi}-1)+\sum_{q\in q^{(O,e)},q\neq\pi}\omega(q)\left(  2\eta_{q}^{\dagger}\eta
_{q}-1\right)  , \label{H(O,e)}
\end{align}%
where
\begin{align}
&q^{(E,o)}=\{-\frac{N-2}{N}\pi,\ldots,-\frac{2}{N}\pi,0,\frac{2}{N}\pi
,\ldots,\frac{N-2}{N}\pi,\pi\},\\
&q^{(E,e)}=\{-\frac{N-1}{N}\pi,\ldots,-\frac{1}{N}\pi,\frac{1}{N}\pi,\ldots
,\frac{N-1}{N}\pi\},\\
&q^{(O,o)}=\{-\frac{N-1}{N}\pi,\ldots,-\frac{2}{N}\pi,0,\frac{2}{N}\pi
,\ldots,\frac{N-1}{N}\pi\},\\
&q^{(O,e)}=\{-\frac{N-2}{N}\pi,\ldots,-\frac{1}{N}\pi,\frac{1}{N}\pi,\ldots
,\frac{N-2}{N}\pi,\pi\}.
\end{align}
So there are four kinds of fermion vacuums that can be expressed in the form of BCS-type wave functions,
\begin{align}
|\phi^{(E/O,e/o)}\rangle &  =\prod_{\substack{q\in q^{(E/O,e/o)},\\(0<q<\pi)}}\left(
u_{q}+\operatorname{i}v_{q}c_{q}^{\dag}c_{-q}^{\dag}\right)  |0\rangle,
\end{align}
above which quasiparticles are created. To restore the exact degrees of freedom of the original spin system, we erase the non-physical states by using the projectors,
\begin{align}
P^{\pm}=\frac{1}{2}\left[  1\pm\prod_{n=1}^{N}\left(  1-2c_{n}^{\dag}c_{n}\right)  \right].
\end{align}
We have
\begin{align}
H(\gamma,h)=P^{+}H^{(E,e)}P^{+}\oplus P^{-}H^{(E,o)}P^{-},
\label{H(2L)}
\end{align}
for $N=2L\in Even$ and
\begin{align}
H(\gamma,h)=P^{+}H^{(O,e)}P^{+}\oplus P^{-}H^{(O,o)}P^{-}.
\label{H(2L+1)}
\end{align}
for $N=2L+1\in Odd$.

In such a tedious but faithful mapping, we clearly see the resemblance and difference between the spin Hamiltonian, Eq. (\ref{XY}), and the fermionic Hamiltonians, Eqs. (\ref{H(E,o)}-\ref{H(O,e)}). For a bipartite lattice, i.e. when $N=2L\in Even$, the ring frustration is absent, so the discrepancy is small and may be neglected. But when $N=2L+1\in Odd$, the system's bulk property is largely changed, because the ring frustration shuffles the ground state and the low-energy excited states \cite{DongJSM,DongPRE}.

In the following , the correlation functions we need in this work are deduced delicately.

\subsection{Transverse Ising ring at its phase transition point}

The transverse Ising model at its phase transition point reads ($\gamma=1, h=1$),
\begin{equation}
H(1,1)=H^{\mathrm{TI}}=\sum_{j=1}^{N}\sigma_{j}^{x}\sigma_{j+1}^{x}-\sum_{j=1}^{N}\sigma_{j}^{z}.
\label{}
\end{equation}

\subsubsection{$N=2L\in Even$}

When $N=2L\in Even$, the ground state is
\begin{align}
|E_{0}^{(E,e)}\rangle=|\phi^{(E,e)}\rangle. \label{}%
\end{align}
and its energy reads
\begin{align}
E_{0}^{(E,e)}=-\sum_{q\in q^{(E,e)}}\omega(q). \label{}%
\end{align}
according to Eq. (\ref{H(2L)}). By introducing the notations, $A_{j}=c_{j}^{\dag}+c_{j}$ and $B_{j}=c_{j}^{\dag}-c_{j}$, applying the Wick's theorem in respect of $|\phi^{\left(E, e\right)  }\rangle$, and retaining the nonzero contractions, $\langle\phi^{\left(E, e\right)  }|B_{l}A_{m}|\phi^{\left(E, e\right)}\rangle= \mathscr{D}_{l-m+1}^{(E,e)}$, the longitudinal correlation function is rewritten in a Toeplitz determinant,
\begin{align}
C_{r,N}=\langle\phi^{\left(E, e\right)  }|B_{j}A_{j+1}\ldots B_{j+r-1}A_{j+r}|\phi^{\left(E, e\right)  }\rangle
=\left\vert
\begin{array}[c]{llll}%
\mathscr{D}_{0}^{(E,e)} & \mathscr{D}_{-1}^{(E,e)} & \cdots &
\mathscr{D}_{-r+1}^{(E,e)}\\
\mathscr{D}_{1}^{(E,e)} & \mathscr{D}_{0}^{(E,e)} & \cdots &
\mathscr{D}_{-r+2}^{(E,e)}\\
\cdots & \cdots & \cdots & \cdots\\
\mathscr{D}_{r-1}^{(E,e)} & \mathscr{D}_{r-2}^{(E,e)} & \cdots &
\mathscr{D}_{0}^{(E,e)}%
\end{array}
\right\vert,
\label{CrEe_TI}%
\end{align}
where
\begin{align}
\mathscr{D}_{n}^{(E,e)}&=\frac{1}{N}\sum_{q\in q^{(E,e)}}D(\mathrm{e}^{\mathrm{i}q})e^{-\mathrm{i}q n},\label{DnEe_TI}\\
D(\mathrm{e}^{\mathrm{i}q})&=\mathrm{e}^{\mathrm{i}q}(1-2u_{q}^{2}+2\mathrm{i}u_{q}v_{q}).
\label{Deiq_TI}
\end{align}
Since (due to $\gamma=1, h=1$)
\begin{align}
D(\mathrm{e}^{\mathrm{i}q})&=\mathrm{i}\sgn(q) \mathrm{e}^{\mathrm{i}q/2},
\label{}
\end{align}
we have
\begin{align}
\mathscr{D}_{r}^{(E,e)}&=-\frac{1}{N}\csc\frac{(1-2r)\pi}{2N}.
\label{}
\end{align}
Thus, for $N=2L\in Even$, we get the abbreviated correlation function in the paper (Eq. (\ref{Cr2L_TI})),
\begin{align}
C_{r,N}=\left(-\frac{1}{N}\right)^{r}\det\Big[\csc\frac{(\mu_j+\nu_k)\pi}{2N}\Big]_{1\leq j,k\leq r},
\label{}%
\end{align}
where $\mu_j=2j+1,~~~\nu_k=-2k$.

\subsubsection{$N=2L+1\in Odd$}

When $N=2L+1\in Odd$, the ground state is
\begin{align}
|E_{0}^{(O,o)}\rangle=c_{0}^{\dag}|\phi^{(O,o)}\rangle,
\label{}%
\end{align}
and its energy is
\begin{align}
E_{0}^{(O,o)}=-\sum_{q\in q^{(O,o)}}\omega(q). \label{}%
\end{align}
according to Eq. (\ref{H(2L+1)}). For the ground state $|E_{0}^{(O,o)}\rangle$, we need to apply the Wick's theorem in respect of $|\phi^{(O,o)}\rangle$,
\begin{align}
C_{r,N}  =\langle\phi^{(O,o)}|c_{0}B_{j}A_{j+1}\ldots B_{j+r-1}A_{j+r}c_{0}^{\dag}|\phi^{(O,o)}\rangle.
\end{align}
We can choose to eliminate the operators $c_{0}$ and $c_{0}^{\dag}$ by nonzero contractions, $\langle\phi^{(O,o)}| c_{0}c_{0}^{\dag}|\phi^{(O,o)}\rangle=1$ and $\langle\phi^{(O,o)}| A_{m}c_{0}^{\dag}|\phi^{(O,o)}\rangle=-\langle\phi^{(O,o)}| B_{m}c_{0}^{\dag}|\phi^{(O,o)}\rangle=\frac{1}{\sqrt{N}}$,
to deduce an expression like
\begin{align}
C_{r,N}&  =\langle\phi^{(O,o)}| B_{j}A_{j+1}\ldots B_{j+r-1}%
A_{j+r}|\phi^{(O,o)}\rangle+\frac{2}{N}\langle\phi^{(O,o)}| B_{j+1}A_{j+2}\ldots B_{j+r-1}%
A_{j+r}|\phi^{(O,o)}\rangle \nonumber\\
&  +\frac{2}{N}\langle\phi^{(O,o)}| A_{j+1}B_{j+1}B_{j+2}A_{j+3}\ldots B_{j+r-1}%
A_{j+r}|\phi^{(O,o)}\rangle +\cdots.   \label{gs}%
\end{align}
Then by nonzero contractions, $\langle\phi^{\left(O,o\right)  }|B_{l}A_{m}|\phi^{\left(O,o\right)}\rangle= \mathscr{D}_{l-m+1}^{(O,o)}$, with
\begin{align}
\mathscr{D}_{n}^{(O,o)}=-\frac{1}{N}+\frac{1}{N}\sum_{_{\substack{q\in q^{\left(O,  o\right)  },q\neq 0}}}%
D(\operatorname{e}^{\operatorname{i}q})\operatorname{e}^{-\operatorname*{i}q n}, \label{scrDn(O,o)}
\end{align}
we can deduce the result as
\begin{align}
C_{r,N}&=
\left\vert
\begin{array}
[c]{cccc}%
\mathscr{D}_{0}^{(O,o)} & \mathscr{D}_{-1}^{(O,o)} & \cdots & \mathscr{D}_{1-r}^{(O,o)}\\
\mathscr{D}_{1}^{(O,o)} & \mathscr{D}_{0}^{(O,o)} & \cdots & \mathscr{D}_{2-r}^{(O,o)}\\
\cdots & \cdots & \cdots & \cdots\\
\mathscr{D}_{r-1}^{(O,o)} & \mathscr{D}_{r-2}^{(O,o)} & \cdots & \mathscr{D}_{0}^{(O,o)}%
\end{array}
\right\vert+\left\vert
\begin{array}
[c]{cccc}%
\frac{2}{N} & \frac{2}{N} & \cdots & \frac{2}{N}\\
\mathscr{D}_{1}^{(O,o)} & \mathscr{D}_{0}^{(O,o)} & \cdots & \mathscr{D}_{2-r}^{(O,o)}\\
\cdots & \cdots & \cdots & \cdots\\
\mathscr{D}_{r-1}^{(O,o)} & \mathscr{D}_{r-2}^{(O,o)} & \cdots & \mathscr{D}_{0}^{(O,o)}%
\end{array}
\right\vert+\cdots+\left\vert
\begin{array}
[c]{cccc}%
\mathscr{D}_{0}^{(O,o)} & \mathscr{D}_{-1}^{(O,o)} & \cdots & \mathscr{D}_{1-r}^{(O,o)}\\
\mathscr{D}_{1}^{(O,o)} & \mathscr{D}_{0}^{(O,o)} & \cdots & \mathscr{D}_{2-r}^{(O,o)}\\
\cdots & \cdots & \cdots & \cdots\\
\frac{2}{N} & \frac{2}{N} & \cdots & \frac{2}{N}%
\end{array}
\right\vert  \nonumber \\
&=\left\vert
\begin{array}
[c]{llll}%
\mathscr{D}_{0}^{(O,o)}+\frac{2}{N} & \mathscr{D}_{-1}^{(O,o)}+\frac{2}{N} & \cdots & \mathscr{D}_{1-r}^{(O,o)}+\frac{2}{N}\\
\mathscr{D}_{1}^{(O,o)}+\frac{2}{N} & \mathscr{D}_{0}^{(O,o)}+\frac{2}{N} & \cdots & \mathscr{D}_{2-r}^{(O,o)}+\frac{2}{N}\\
\vdots & \vdots & \vdots & \vdots\\
\mathscr{D}_{r-1}^{(O,o)}+\frac{2}{N} & \mathscr{D}_{r-2}^{(O,o)}+\frac{2}{N} & \cdots & \mathscr{D}_{0}^{(O,o)}+\frac{2}{N}%
\end{array}
\right\vert .\label{corgs}
\end{align}
And since (due to $\gamma=1,h=1$)
\begin{equation}
D(\mathrm{e}^{\mathrm{i}q})=\mathrm{i}\sgn(q) \mathrm{e}^{\mathrm{i}q/2},
\label{Dq}
\end{equation}
we have
\begin{equation}
\mathscr{D}_{r}^{(O,o)}=-\frac{1}{N}-\frac{1}{N}\cot\frac{\left(1-2r\right)\pi}{2N}.
\end{equation}
Thus for $N=2L+1\in Odd$, we get the Toeplitz determinant representation of the correlation function in the paper (Eq. (\ref{Cr2L1_TI}))
\begin{align}
C_{r,N}=\left(\frac{1}{N}\right)^{r}\det\Big[1-\cot\frac{\left(\mu_j+\nu_k\right)\pi}{2N}\Big]_{0\leq j,k\leq r-1}.
\label{}%
\end{align}

\subsection{Isotropic $XY$ ring}

The isotropic $XY$ model reads ($\gamma=0, h=0$),
\begin{equation}
H(0,0)=H^{XY}=\frac{1}{2}\sum_{j=1}^{N}\left(\sigma_{j}^{x}\sigma_{j+1}^{x}+\sigma_{j}^{y}\sigma_{j+1}^{y}\right).
\label{}
\end{equation}
The situation in $XY$ ring is more delicate than that in transverse Ising ring. The solutions for the ground state ought to be put into 4 categories:

(\emph{a}) For $N=4K~(K=1,2,3,\cdots)$, the ground state is unique, which reads $|\phi^{(E,e)}\rangle$.

(\emph{b}) For $N=4K+2$, the ground state is unique, which reads $c_{\pi}^{\dag}|\phi^{(E,o)}\rangle$.

(\emph{c}) For $N=4K+1$, the ground states are four-fold degenerate due to the presence of ring frustration. Two of them come from the odd channel,
\begin{equation}
|E_{\pm Q_{o}}^{(O,o)}\rangle = \eta_{\pm Q_{o}}^{\dagger}|\phi^{(O,o)}\rangle,
\label{Qostates}
\end{equation}
and two of them from the even channel,
\begin{equation}
|E_{\pm Q_{e}}^{(O,e)}\rangle = \eta_{\pm Q_{e}}^{\dagger} c_{\pi}^{\dagger}|\phi^{(O,e)}\rangle.
\label{Qestates}
\end{equation}
where the characteristic wave vectors are
\begin{align}
Q_{o} = \frac{N-1}{2N}\pi,~Q_{e} = \frac{N+1}{2N}\pi.
\end{align}

(\emph{d}) For $N=4K+3$, the ground states are four-fold degenerate due to the presence of ring frustration. They are also expressed by Eqs. (\ref{Qostates}) and (\ref{Qestates}), but the characteristic wave vectors swaps
\begin{align}
Q_{o}= \frac{N+1}{2N}\pi,~Q_{e} = \frac{N-1}{2N}\pi.
\end{align}

Let us demonstrate their correlation functions in Toeplitz determinant representation one by one.

\subsubsection{$N=4K\in Even$}
In this case, because the ground state is $|\phi^{(E,e)}\rangle$, the correlation function shares the same expressions as that in Eqs. (\ref{CrEe_TI})-(\ref{Deiq_TI}), but the elements are different and read,
\begin{align}
\mathscr{D}_{r}^{(E,e)}&=\left\{
\begin{array}[c]{ll}%
0,&(r=1);\\
-\frac{2}{N}\csc\frac{\pi (r-1)}{N}\sin\frac{\pi (r-1)}{2},&(\mathrm{other~} r),%
\end{array}
\right.
\label{}
\end{align}
since now we have
\begin{align}
D(\mathrm{e}^{\mathrm{i}q})&=-\sgn(\cos q) \mathrm{e}^{\mathrm{i}q}.
\end{align}

\subsubsection{$N=4K+2\in Even$}

In this case, the ground state is $c_{\pi}^{\dag}|\phi^{(E,o)}\rangle$. We need to apply the Wick's theorem in respect of $|\phi^{(E,o)}\rangle$,
\begin{align}
C_{r,N} &  =\langle\phi^{(E,o)}|c_{\pi}B_{j}A_{j+1}\ldots B_{j+r-1}A_{j+r}c_{\pi}^{\dag}|\phi^{(E,o)}\rangle.
\end{align}
We can choose to eliminate the operators $c_{\pi}$ and $c_{\pi}^{\dag}$ first by using nonzero contractions, $\langle\phi^{(E,o)}| c_{\pi}c_{\pi}^{\dag}|\phi^{(E,o)}\rangle=1$ and $\langle\phi^{(E,o)}| A_{m}c_{\pi}^{\dag}|\phi^{(E,o)}\rangle=-\langle\phi^{(E,o)}| B_{m}c_{\pi}^{\dag}|\phi^{(E,o)}\rangle=\frac{(-1)^m}{\sqrt{N}}$,
to get an expression like
\begin{align}
C_{r,N} &  =\langle\phi^{(E,o)}| B_{j}A_{j+1}\ldots B_{j+r-1}%
A_{j+r}|\phi^{(E,o)}\rangle-\frac{2}{N}\langle\phi^{(E,o)}| B_{j+1}A_{j+2}\ldots B_{j+r-1}%
A_{j+r}|\phi^{(E,o)}\rangle \nonumber\\
& -\frac{2}{N}\langle\phi^{(E,o)}| A_{j+1}B_{j+1}B_{j+2}A_{j+3}\ldots B_{j+r-1}%
A_{j+r}|\phi^{(E,o)}\rangle +\cdots.   \label{gs}%
\end{align}
Then by nonzero contractions, $\langle\phi^{\left(E,o\right)  }|B_{l}A_{m}|\phi^{\left(E,o\right)}\rangle= \mathscr{D}_{l-m+1}^{(E,o)}$, with
\begin{align}
&\mathscr{D}_{n}^{(E,o)}=-\frac{1}{N}+\frac{1}{N}\sum_{_{\substack{q\in q^{\left(E,o\right)  },q\neq \pi}}}%
D(\operatorname{e}^{\operatorname{i}q})\operatorname{e}^{-\operatorname*{i}q n},\\
&D(\operatorname{e}^{\operatorname{i}q}) =\operatorname{e}^{\operatorname*{i}q}(1-2u_{q}^{2}+2\operatorname{i}u_{q}v_{q}),
\end{align}
we get
\begin{align}
C_{r,N}=\left\vert
\begin{array}
[c]{llll}%
\mathscr{D}_{0}^{(E,o)}-\frac{2}{N} & \mathscr{D}_{-1}^{(E,o)}-\frac{2}{N}\mathrm{e}^{-\mathrm{i}\pi} & \cdots & \mathscr{D}_{-(r-1)}^{(E,o)}-\frac{2}{N}\mathrm{e}^{-\mathrm{i}(r-1)\pi}\\
\mathscr{D}_{1}^{(E,o)}-\frac{2}{N}\mathrm{e}^{\mathrm{i}\pi} & \mathscr{D}_{0}^{(E,o)}-\frac{2}{N} & \cdots & \mathscr{D}_{-(r-2)}^{(E,o)}-\frac{2}{N}\mathrm{e}^{-\mathrm{i}(r-2)\pi}\\
\vdots & \vdots & \vdots & \vdots\\
\mathscr{D}_{r-1}^{(E,o)}-\frac{2}{N}\mathrm{e}^{\mathrm{i}(r-1)\pi} & \mathscr{D}_{r-2}^{(E,o)}-\frac{2}{N}\mathrm{e}^{-\mathrm{i}(r-2)\pi} & \cdots & \mathscr{D}_{0}^{(E,o)}-\frac{2}{N}%
\end{array}
\right\vert .\label{}
\end{align}

For isotropic $XY$ model ($\gamma=0,h=0$), we have
\begin{equation}
D(\mathrm{e}^{\mathrm{i}q})=-\sgn(\cos q) \mathrm{e}^{\mathrm{i}q},
\label{}
\end{equation}
so we get
\begin{align}
\mathscr{D}_{n}^{(E,o)}&=\left\{
\begin{array}[c]{ll}%
-\frac{2}{N},&(n=1);\\
-\frac{2}{N}-\frac{4}{N}\csc\frac{\pi (n-1)}{N}\sin\frac{\pi (n-1)}{2}\left[\sin\frac{(N-2)(n-1)\pi}{4N}\right]^{2},&(\mathrm{other~} n).%
\end{array}
\right.
\label{}
\end{align}

\subsubsection{$N=4K+1\in Odd$}

The ground states are of four degeneracy. For simplicity and without loss of generality, let us choose the state $|E_{Q_{o}}^{(O,o)}\rangle=\eta_{Q_{o}}^{\dagger}|\phi^{(O,o)}\rangle$. The starting point is%
\begin{equation}
C_{r,N} =\langle\phi^{(O,o)}|\eta
_{Q_{o}}B_{j}A_{j+1}\ldots B_{j+r-1}A_{j+r}\eta_{Q_{o}}^{\dagger}|\phi^{(O,o)}\rangle.
\label{def_CxxEko}
\end{equation}
Likewise, the strategy is to eliminate the operators $\eta_{Q_{o}}$ and $\eta_{Q_{o}}^{\dag}$ first. Except for $\langle\phi^{(O,o)}| \eta_{Q_{o}} \eta_{Q_{o}}^{\dag}|\phi^{(O,o)}\rangle=1$, we find the combined nonzero contractions are very useful%
\begin{align}
\langle\phi^{(O,o)}|\eta_{Q_{o}}B_{l}|\phi^{(O,o)}\rangle&\langle\phi^{(O,o)}| A_{m}\eta_{Q_{o}}^{\dagger}|\phi^{(O,o)}\rangle =
\frac{\beta_{Q_{o}}}{N}\operatorname{e}^{\operatorname*{i}Q_{o}\,\left(  l-m+1\right)},\\
&\beta_{Q_{o}}=-D(\operatorname{e}^{-\operatorname*{i}Q_{o}}).
\end{align}
So we could write down
\begin{align}
2 C_{r,N}=  &\Big[\langle\phi^{(O,o)}|
B_{j}A_{j+1}\ldots B_{j+r-1}A_{j+r}|\phi^{(O,o)}\rangle
+\frac{2\beta_{Q_{o}}}{N}\langle\phi^{(O,o)}| B_{j+1}A_{j+2}\ldots
B_{j+r-1}A_{j+r}|\phi^{(O,o)}\rangle \nonumber\\
&~~ +\frac{2\beta_{Q_{o}}\operatorname{e}^{-\operatorname*{i}Q_{o}}}{N}\langle\phi^{(O,o)}|
A_{j+1}B_{j+1}B_{j+2}A_{j+3}\ldots B_{j+r-1}A_{j+r}|\phi^{(O,o)}\rangle
+\cdots \Big]\nonumber\\
+& \Big[\langle\phi^{(O,o)}| B_{j}A_{j+1}\ldots B_{j+r-1}%
A_{j+r}|\phi^{(O,o)}\rangle +\frac{2\beta_{-Q_{o}}}{N}\langle\phi^{(O,o)}| B_{j+1}A_{j+2}%
\ldots B_{j+r-1}A_{j+r}|\phi^{(O,o)}\rangle \nonumber\\
&~~  +\frac{2\beta_{-Q_{o}}\operatorname{e}^{\operatorname*{i}Q_{o}}}{N}\langle\phi^{(O,o)}|
A_{j+1}B_{j+1}B_{j+2}A_{j+3}\ldots B_{j+r-1}A_{j+r}|\phi^{(O,o)}\rangle +\cdots \Big]~.
\label{}
\end{align}
The terms are grouped into two square brackets. Thus the correlation function can be represented by the sum of two Toeplitz determinants,
\begin{align}
C_{r,N} &=\frac{1}{2}\left[
\Gamma^{(O,o)}\left(  r,N,\beta_{Q_{o}},\operatorname{e}^{\operatorname{i}Q_{o}}\right)
+\Gamma^{(O,o)}\left(  r,N,\beta_{-Q_{o}},\operatorname{e}^{-\operatorname{i}%
Q_{o}}\right)  \right]\nonumber\\
&=\Re\left[\Gamma^{(O,o)}\left(  r,N,\beta_{Q_{o}},\operatorname{e}^{\operatorname{i}Q_{o}}\right)\right]  ,
\label{}%
\end{align}
where $\Re[~]$ means taking the real part of the number and the determinant $\Gamma^{(O,o)}(r,N,\beta_{Q_{o}},\operatorname{e}^{\operatorname*{i}Q_{o}})$ reads
\begin{equation}
\Gamma^{(O,o)}(r,N,\beta_{Q_{o}},\operatorname{e}^{\operatorname*{i}Q_{o}})=\left\vert
\begin{array}
[c]{llll}%
\mathscr{D}_{0}^{(O,o)}+\frac{2\beta_{Q_{o}}}{N} & \mathscr{D}_{-1}^{(O,o)}+\frac{2\beta_{Q_{o}}}{N}\operatorname{e}%
^{-\operatorname*{i}Q_{o}} & \cdots & \mathscr{D}_{1-r}^{(O,o)}+\frac{2\beta_{Q_{o}}}{N}\operatorname{e}%
^{\operatorname*{i}(1-r)Q_{o}}\\
\mathscr{D}_{1}^{(O,o)}+\frac{2\beta_{Q_{o}}}{N}\operatorname{e}^{\operatorname*{i}Q_{o}} & \mathscr{D}_{0}^{(O,o)}%
+\frac{2\beta_{Q_{o}}}{N} & \cdots & \mathscr{D}_{2-r}^{(O,o)}+\frac{2\beta_{Q_{o}}}{N}\operatorname{e}%
^{\operatorname*{i}(2-r)Q_{o}}\\
\vdots & \vdots & \vdots & \vdots\\
\mathscr{D}_{r-1}^{(O,o)}+\frac{2\beta_{Q_{o}}}{N}\operatorname{e}^{\operatorname*{i}(r-1)Q_{o}} &
\mathscr{D}_{r-2}^{(O,o)}+\frac{2\beta_{Q_{o}}}{N}\operatorname{e}^{\operatorname*{i}(r-2)Q_{o}} &
\cdots & \mathscr{D}_{0}^{(O,o)}+\frac{2\beta_{Q_{o}}}{N}%
\end{array}
\right\vert ,
\label{corEko1}
\end{equation}
with
\begin{equation}
\beta_{Q_{o}}=-D(\operatorname{e}^{-\operatorname*{i}Q_{o}})=\sgn(\cos Q_{o})\operatorname{e}^{-\operatorname*{i}Q_{o}}.
\end{equation}
$\mathscr{D}_{n}^{(O,o)}$ is defined in Eq. (\ref{scrDn(O,o)}) and we have
\begin{align}
\mathscr{D}_{n}^{(O,o)}=\left\{
\begin{array}[c]{ll}%
-\frac{1}{N},~~&(n = 1);\\
-\frac{2}{N}\csc\frac{(n-1)\pi}{N}\sin\frac{(N+1)(n-1)\pi}{2N}, ~&(\mathrm{other~}n).%
\end{array}
\right.
\label{}
\end{align}

\subsubsection{$N=4K+3\in Odd$}

We also choose the state $|E_{Q_{o}}^{(O,o)}\rangle=\eta_{Q_{o}}^{\dagger}|\phi^{(O,o)}\rangle$. It turns out the deduction is almost the same as that for $N=4K+1$, except for the final expression for $\mathscr{D}_{r-1}^{(O,o)}$,
\begin{align}
\mathscr{D}_{n}^{(O,o)}=\left\{
\begin{array}[c]{ll}%
\frac{1}{N},~~&(n = 1);\\
-\frac{2}{N}\csc\frac{(n-1)\pi}{N}\sin\frac{(N-1)(n-1)\pi}{2N}, ~&(\mathrm{other~}n).%
\end{array}
\right.
\label{}
\end{align}

\section{Asymptotic analysis of Eq. (\ref{SrN1})}
\label{secSrN}

For $N=2L+1$, we have an exact identity (for $N=2L$, it is approximate),
\begin{align}
\frac{1}{N \sin\theta}=\prod_{1\leq m\leq L}\left(\cos^{2}\theta-\cot^{2}m\theta\sin^{2}\theta\right),
\label{identity_app}
\end{align}
so we find
\begin{align}
\ln S_{r,N}&=U_{r,N}+V_{r,N},\label{lnSrN_app}\\
U_{r,N}&=\sum_{m=1}^{r-1}m \ln\left(\cos^{2}\theta-\cot^{2}m\theta\sin^{2}\theta\right),\label{UrN}\\
V_{r,N}&=r\sum_{m=r}^{(N-1)/2}\ln\left(\cos^{2}\theta-\cot^{2}m\theta\sin^{2}\theta\right).\label{VrN}
\end{align}

Introducing the Taylor expansion,
\begin{align}
\ln\left(\cos^{2}\theta-\cot^{2}m\theta\sin^{2}\theta\right)=
\ln(1-\frac{1}{4m^2})-\frac{1}{3}\theta^2-\frac{1+24m^2}{90}\theta^4-\frac{2(1+60m^2+240m^4)}{2835}\theta^6-\cdots,
\label{expansion}
\end{align}
substituting it into Eq. (\ref{UrN}) and accomplishing the summation, we get
\begin{align}
U_{r,N}=g_{1}(\alpha)+\sum_{m=1}^{r-1}m \ln\left(1-\frac{1}{4m^2}\right)+O(\frac{1}{N}),\label{UrN2}
\end{align}
where $g_1(\alpha)$ is a convergent series,
\begin{align}
g_{1}(\alpha)=-\frac{\pi^2\alpha^2}{24}-\frac{\pi^4\alpha^4}{240}-\frac{\pi^6\alpha^6}{2268}-
\frac{\pi^8\alpha^8}{21600}-\frac{\pi^{10}\alpha^{10}}{207900}
-\frac{691\pi^{12}\alpha^{12}}{1393119000}-\frac{\pi^{14}\alpha^{14}}{19646550}
-\frac{3617\pi^{16}\alpha^{16}}{694702008000}-\cdots,
\end{align}
the second term turns out to be \cite{Wu,McCoy}
\begin{align}
\sum_{m=1}^{r-1}m \ln\left(1-\frac{1}{4m^2}\right)\approx\frac{1}{4}-\frac{1}{4}\ln r+\ln b_{1},
\end{align}
where $b_{1} =\mathrm{e}^{1/4}2^{1/12}A^{-3}\approx 0.645002448$, $A=1.28242713$ is the Glaisher constant.

Likewise, by substituting Eq. (\ref{expansion}) into (\ref{VrN}) and accomplishing the summation, we get
\begin{align}
V_{r,N}=g_{2}(\alpha)+r\sum_{m=r}^{(N-1)/2} \ln\left(1-\frac{1}{4m^2}\right)+O(\frac{1}{N}),\label{VrN2}
\end{align}
where $g_2(\alpha)$ is another convergent series,
\begin{align}
g_{2}(\alpha)=&-\frac{\pi^2}{24}\alpha(1-2\alpha)-\frac{\pi^4}{1440}\alpha[1-(2\alpha)^3]
-\frac{\pi^6}{60480}\alpha[1-(2\alpha)^5]-\frac{\pi^8}{2419200}\alpha[1-(2\alpha)^7]-\frac{\pi^{10}}{95800320}\alpha[1-(2\alpha)^9]\nonumber\\
&-\frac{691\pi^{12}}{2615348736000}\alpha[1-(2\alpha)^{11}]
-\frac{\pi^{14}}{149448499200}\alpha[1-(2\alpha)^{13}]-\frac{3617\pi^{16}}{21341245685760000}\alpha[1-(2\alpha)^{15}]-\cdots,
\end{align}
and, as the leading order, the second term is tackled as
\begin{align}
r\sum_{m=r}^{(N-1)/2} \ln\left(1-\frac{1}{4m^2}\right)\approx-\frac{r}{4}\int_{r}^{(N-1)/2}\frac{1}{m^2}\mathrm{d}m\approx-\frac{1-2\alpha}{4}.
\end{align}
Compared with the traditional result, $-\frac{1}{4}$ \cite{Wu,McCoy}, our result show that an extra factor $\frac{\alpha}{2}$ is dropped in the leading order.
At last, we finish the analysis by summing up all the essential terms and writing down
\begin{align}
S(\alpha)\equiv\lim_{N\rightarrow\infty}S_{r,N}=\frac{b_{1}}{r^{1/4}}\mathrm{e}^{h(\alpha)},
\end{align}
where we have defined
\begin{align}
h(\alpha)=\frac{\alpha}{2}+g_{1}(\alpha)+g_{2}(\alpha).
\label{h_app}
\end{align}

\section{Derivation of Eqs. (\ref{Pn}) and (\ref{Qn})}
\label{triCauchy}

We will first prove the identity Eq. (\ref{Qn}) by mathematical recursion. For convenience, we repeat  Eq.(\ref{Qn}) here
\be
D_n(\phi)=\det\Big[\frac{\cos(a_i+b_j+\phi)}{\sin(a_i+b_j)}\Big]_{1\leq i,j\leq n}
=\frac{\prod_{1\leq i<j\leq n}\sin(a_i-a_j)\sin(b_i-b_j)}{\prod_{1\leq i,j\leq n}\sin(a_i+b_j)}
\cos\Big[\sum_{i=1}^n(a_i+b_i)+\phi\Big]\cos^{n-1}\phi.  \label{Dn}
\ee
Write out $D_n(\phi)$ explicitly as
\be
D_n(\phi)=\left|
\begin{array}{cccc}
\dfrac{\cos(a_1+b_1+\phi)}{\sin(a_1+b_1)} & \dfrac{\cos(a_1+b_2+\phi)}{\sin(a_1+b_2)}
& \cdots & \dfrac{\cos(a_1+b_n+\phi)}{\sin(a_1+b_n)} \\
\dfrac{\cos(a_2+b_1+\phi)}{\sin(a_2+b_1)} & \dfrac{\cos(a_2+b_2+\phi)}{\sin(a_2+b_2)}
& \cdots & \dfrac{\cos(a_2+b_n+\phi)}{\sin(a_2+b_n)} \\
\vdots & \vdots & \ddots & \vdots \\
\dfrac{\cos(a_n+b_1+\phi)}{\sin(a_n+b_1)} & \dfrac{\cos(a_n+b_2+\phi)}{\sin(a_n+b_2)}
& \cdots & \dfrac{\cos(a_n+b_n+\phi)}{\sin(a_n+b_n)}
      \end{array}
    \right|
\ee
Subtract the last column from all previous columns, we find
\be
&&D_n(\phi)=\left|
\begin{array}{cccc}
\dfrac{-\cos\phi\sin(b_1-b_n)}{\sin(a_1+b_1)\sin(a_1+b_n)}
& \dfrac{-\cos\phi\sin(b_2-b_n)}{\sin(a_1+b_2)\sin(a_1+b_n)}
& \cdots & \dfrac{\cos(a_1+b_n+\phi)}{\sin(a_1+b_n)} \\
\dfrac{-\cos\phi\sin(b_1-b_n)}{\sin(a_2+b_1)\sin(a_2+b_n)}
&\dfrac{-\cos\phi\sin(b_1-b_n)}{\sin(a_2+b_2)\sin(a_2+b_n)}
& \cdots & \dfrac{\cos(a_2+b_n+\phi)}{\sin(a_2+b_n)} \\
\vdots & \vdots & \ddots & \vdots \\
\dfrac{-\cos\phi\sin(b_1-b_n)}{\sin(a_n+b_1)\sin(a_n+b_n)}
& \dfrac{-\cos\phi\sin(b_2-b_n)}{\sin(a_n+b_2)\sin(a_n+b_n)}
& \cdots & \dfrac{\cos(a_n+b_n+\phi)}{\sin(a_n+b_n)}
      \end{array}
    \right|\nonumber\\
&&=(-\cos\phi)^{n-1}\frac{\prod_{i=1}^{n-1}\sin(b_i-b_n)}{\prod_{j=1}^n\sin(a_j+b_n)}
\left|
\begin{array}{cccc}
\dfrac{1}{\sin(a_1+b_1)} & \dfrac{1}{\sin(a_1+b_2)} & \cdots & \cos(a_1+b_n+\phi) \\
\dfrac{1}{\sin(a_2+b_1)} & \dfrac{1}{\sin(a_2+b_2)} & \cdots & \cos(a_2+b_n+\phi) \\
\vdots & \vdots & \ddots & \vdots \\
\dfrac{1}{\sin(a_n+b_1)} & \dfrac{1}{\sin(a_n+b_2)} & \cdots & \cos(a_n+b_n+\phi)
      \end{array}
\right|\nonumber\\
\ee
In the above determinant, multiply the last row by $\frac{-\cos(a_i+b_n+\phi)}{\cos(a_n+b_n+\phi)}$ and add to the $i$th row for $i=1,\cdots, n-1$. For the element at $(i,j)$, we have
\be
\frac{1}{\sin(a_i+b_j)}+\frac{1}{\sin(a_n+b_j)}\frac{-\cos(a_i+b_n+\phi)}{\cos(a_n+b_n+\phi)}
=\frac{-\cos(a_i+b_j+a_n+b_n+\phi)\sin(a_i-a_n)}{\sin(a_i+b_j)\sin(a_n+b_j)\cos(a_n+b_n+\phi)}
\ee
Make use of this identity, extract common factor for each row and Column and define $\phi'=\phi+a_n+b_n$, we find
\be
&&D_n(\phi)=(\frac{\cos\phi}{\cos\phi'})^{n-1}\frac{\prod_{i=1}^{n-1}\sin(b_i-b_n)\sin(a_i-a_n)}
{\prod_{j=1}^n\sin(a_j+b_n)\prod_{j=1}^{n-1}\sin(a_n+b_j)}\left|
\begin{array}{cccc}
\dfrac{\cos(a_1+b_1+\phi')}{\sin(a_1+b_1)} & \dfrac{\cos(a_1+b_2+\phi')}{\sin(a_1+b_2)} & \cdots & 0\\
\dfrac{\cos(a_2+b_1+\phi')}{\sin(a_2+b_1)} & \dfrac{\cos(a_2+b_2+\phi')}{\sin(a_2+b_2)} & \cdots & 0\\
\vdots & \vdots & \ddots & \vdots \\
1 & 1 & \cdots & \cos\phi'
\end{array}
\right|\nonumber\\
&&=\frac{\cos^{n-1}\phi}{\cos^{n-2}\phi'}\frac{\prod_{i=1}^{n-1}\sin(b_i-b_n)\sin(a_i-a_n)}
{\prod_{j=1}^n\sin(a_j+b_n)\prod_{j=1}^{n-1}\sin(a_n+b_j)}D_{n-1}(\phi')\label{recur}
\ee
By mathematical recursion assumption, we have
\be
D_{n-1}(\phi')=\frac{\prod_{1\leq i<j\leq n-1}\sin(a_i-a_j)\sin(b_i-b_j)}{\prod_{1\leq i,j\leq n-1}\sin(a_i+b_j)}
\cos\Big[\sum_{i=1}^{n-1}(a_i+b_i)+\phi'\Big]\cos^{n-1}\phi'\label{Dn1}
\ee
Plug Eq.(\ref{Dn1}) into Eq.(\ref{recur}), we get back Eq.(\ref{Dn}), which finishes the proof of Eq.(\ref{Qn}).

We then prove the identity Eq. (\ref{Pn}) again by mathematical recursion. For convenience, we repeat  Eq.(\ref{Pn}) here
\be
A_n=\det\Big[\frac{1}{\sin(a_i+b_j)}\Big]_{1\leq i,j\leq n}
=\frac{\prod_{1\leq i<j\leq n}\sin(a_i-a_j)\sin(b_i-b_j)}{\prod_{1\leq i,j\leq n}\sin(a_i+b_j)}
\label{An}
\ee
Write out $A_n$ explicitly as
\be
A_n=\left|
\begin{array}{cccc}
\dfrac{1}{\sin(a_1+b_1)} & \dfrac{1}{\sin(a_1+b_2)}
& \cdots & \dfrac{1}{\sin(a_1+b_n)} \\
\dfrac{1}{\sin(a_2+b_1)} & \dfrac{1}{\sin(a_2+b_2)}
& \cdots & \dfrac{1}{\sin(a_2+b_n)} \\
\vdots & \vdots & \ddots & \vdots \\
\dfrac{1}{\sin(a_n+b_1)} & \dfrac{1}{\sin(a_n+b_2)}
& \cdots & \dfrac{1}{\sin(a_n+b_n)}
      \end{array}
    \right|
\ee
In the above determinant, multiply the last column by $\frac{-\sin b_n}{\sin b_i}$ and add to the $i$th column for $i=1,\cdots, n-1$. For the element at $(i,j)$, we have
\be
\frac{1}{\sin(a_i+b_j)}+\frac{1}{\sin(a_i+b_n)}\frac{-\sin(b_n)}{\sin(b_j)}
=\frac{\sin a_i\sin(b_j-b_n)}{\sin(a_i+b_j)\sin(a_i+b_n)\sin b_j}
\ee
Make use of this identity, extract common factor for each row and Column, we find
\be
A_n=\frac{\prod_{i=1}^{n-1}\sin(b_i-b_n)}{\prod_{j=1}^n\sin(a_j+b_n)}
\left|
\begin{array}{cccc}
\dfrac{\sin a_1}{\sin b_1\sin(a_1+b_1)} & \dfrac{\sin a_1}{\sin b_2\sin(a_1+b_2)} & \cdots & 1 \\
\dfrac{\sin a_2}{\sin b_1\sin(a_2+b_1)} & \dfrac{\sin a_2}{\sin b_2\sin(a_2+b_2)} & \cdots & 1 \\
\vdots & \vdots & \ddots & \vdots \\
\dfrac{\sin a_n}{\sin b_1\sin(a_n+b_1)} & \dfrac{\sin a_n}{\sin b_2\sin(a_n+b_2)} & \cdots & 1
      \end{array}
\right|\nonumber\\
\ee
Subtract the last row from all previous rows and extract common factors, we find
\be
A_n&=&\frac{\prod_{i=1}^{n-1}\sin(b_i-b_n)\sin(a_i-a_n)}{\prod_{j=1}^n\sin(a_j+b_n)\prod_{j=1}^{n-1}\sin(a_n+b_j)}
\left|
\begin{array}{cccc}
\dfrac{1}{\sin(a_1+b_1)} & \dfrac{1}{\sin(a_1+b_2)} & \cdots & 0 \\
\dfrac{1}{\sin(a_2+b_1)} & \dfrac{1}{\sin(a_2+b_2)} & \cdots & 0 \\
\vdots & \vdots & \ddots & \vdots \\
\dfrac{\sin a_n}{\sin b_1\sin(a_n+b_1)} & \dfrac{\sin a_n}{\sin b_2\sin(a_n+b_2)} & \cdots & 1
      \end{array}
\right|\nonumber\\
&=&\frac{\prod_{i=1}^{n-1}\sin(b_i-b_n)\sin(a_i-a_n)}{\prod_{j=1}^n\sin(a_j+b_n)\prod_{j=1}^{n-1}\sin(a_n+b_j)}A_{n-1}
\label{recur1}
\ee
By mathematical recursion assumption, we have
\be
A_{n-1}=\frac{\prod_{1\leq i<j\leq n-1}\sin(a_i-a_j)\sin(b_i-b_j)}{\prod_{1\leq i,j\leq n-1}\sin(a_i+b_j)}
\label{An1}
\ee
Plug Eq.(\ref{An1}) into Eq.(\ref{recur1}), we get back Eq.(\ref{An}), which finishes the proof of Eq.(\ref{Pn}).

\section{Spin-1/2 Heisenberg chain solved by Bethe Ansatz}

In this section, we turn to the isotropic spin-1/2 Heisenberg model with ring frustration. The Hamiltonian is
\be
H^H=J\sum_{i=1}^N(S^x_iS^x_{i+1}+S^y_iS^y_{i+1}+S^z_iS^z_{i+1})
\ee
Similar to the $XY$ model, here we impose the periodic boundary condition $S^a_{N+1}=S^a_1$ for $a=x,y,z$ and only consider the anti-ferromagnetic interacting $J>0$ with odd number of sites.

It is well known that Heisenberg model can be exactly solved by Bethe ansatz. Here we only present the results we need for the calculations of ground state correlation functions, detailed derivation can be found in \cite{Takahashi,FranchiniBook}. First the number of down spins $M$ is conserved in the Heisenberg model, thus we can diagonalize the Hamiltonian in each sub-Hilbert space with fixed number of down spins. Since $J>0$ and $N$ is odd, the ground states occur in the the subspace with $M=(N-1)/2$ or $M=(N+1)/2$. These two subspace can be mapped to each other by a spin flip of all spins. Therefore we only need to consider the case with $M=(N-1)/2$.

The eigenfunction can be expressed as
\be
|\psi\rangle=\sum_{n_1,\cdots,n_M}f(n_1,n_2,\cdots,n_M)|n_1,n_2,\cdots,n_M\rangle
\ee
Here $|n_1,n_2,\cdots,n_M\rangle$ denotes the spin state with $M$ down spins located at lattice site $n_1$ to $n_M$ while all other sites are spin up. The basic idea of Bethe ansatz is to assume that the eigenfunction can be written as a superposition of plane waves as
\be
f(n_1,n_2,\cdots,n_M)=\sum_P A(P)\prod_{j=1}^M\Big(\frac{x_{Pj}+i}{x_{Pj}-i}\Big)^{n_j}
\label{fn}
\ee
Here $x_j$ for $j=1,\cdots, M$ are usually called Bethe roots, which will be determined later. $P$ denotes the permutation of Bethe roots
The requirement that $|\psi\rangle$ to be an eigenstate of $H^H$ determines the amplitude $A(P)$ in terms of Bethe roots as follows
\be
A(P)=A_0\epsilon(P)\prod_{j<l}(x_{Pj}-x_{Pl}+2i)
\label{AP}
\ee
Here $\epsilon(P)=1$ if $P$ is an even permutation and $\epsilon(P)=-1$ if $P$ is an odd permutation. $A_0$ is overall normalization factor.

The periodic boundary condition gives rise the following Bethe equations
\be
\Big(\frac{x_j+i}{x_j-i}\Big)^N=\prod_{l\neq j}\Big(\frac{x_j-x_l+2i}{x_j-x_l-2i}\Big)
\ee
which determines the Bethe roots. To solve the above equation, it is more convenient to take logarithm of both sides and find
\be
2N\arctan x_j=2\pi I_j+2\sum_{l=1}^M\arctan\frac{x_j-x_l}{2}\label{BAE}
\ee
where $I_j$ for $j=1,\cdots, M$ are an integers if $N-M$ is odd, and are half-odd integer if $N-M$ is even.
All eigenfunctions of $H^H$ can be obtained by solving Eq.(\ref{BAE}) with all possible choices of different sets of $I_j$. Substitute the solved Bethe roots into Eq.(\ref{AP}) and Eq.(\ref{fn}), then the exact eigenfunction is obtained.

The ground state corresponds the most symmetric and uniform distribution of $I_j$. For odd $N$, if $M=(N-1)/2$ is even, we can take the following two choices for $I_j$
\be
\{I_j\}=\{-\frac{M}2+1,\cdots,-1,0,1,\cdots,\frac{M}2\}\nonumber\\
\{I_j\}=\{-\frac{M}2,\cdots,-1,0,1,\cdots,\frac{M}2-1\}\nonumber
\ee
If $M$ is odd, we take $I_j$ as
\be
\{I_j\}=\{-\frac{M}2+1,\cdots,-\frac32,-\frac12,\frac12,\frac32,\cdots,\frac{M}2\}\nonumber\\
\{I_j\}=\{-\frac{M}2,\cdots,-\frac32,-\frac12,\frac12,\frac32,\cdots,\frac{M}2-1\}\nonumber
\ee
Note that both cases there is a hole either located at the left end or the right end. One can verifies that these two sets of $I_j$ gives the two degenerate ground states. Recall that we can flip all spins to find another two degenerate states with the same energy in the subspace with $M=(N+1)/2$. Therefore the total ground state degeneracy of anti-ferromagnetic Heisenberg model with odd number of sites is 4, the same as the isotropic $XY$ model.

By taking one of the above sets of $I_j$, we numerically solve Bethe equations to obtain the Bethe roots for the ground state. This computation can be done for a chain of hundreds lattice sites. Summing all the permutations numerically, we find the eigenfunction $f(n_1,\cdots,n_M)$. The ground state correlation is given by
\begin{align}
C^{H}&(r,N)=\langle S^z_1S^z_r\rangle \nonumber\\
&=\frac{\sum\prod_{j=1}^M
(-1)^{\delta_{n_j1}}(-1)^{\delta_{n_j}r}|f(n_1,\cdots,n_M)|^2}
{4\sum|f(n_1,\cdots,n_M)|^2}
\end{align}
Because the number of permutation increases very fast,  we have only computed the correlations $C_{r,N}$ for lattices with number of spins from $N=10$ to $21$.

\end{widetext}

\end{document}